\documentclass[12pt]{article}

\usepackage{hyperref}
\usepackage[inline,shortlabels]{enumitem}
\usepackage{float}
\usepackage{mathtools}
\usepackage{bbm}
\usepackage{graphicx}
\usepackage[square,numbers]{natbib}
\usepackage[margin=1in]{geometry}
\usepackage{tikz}
\usepackage[english]{babel}
\usepackage{longtable}
\usepackage{color}
\usepackage{amssymb,amsmath,amsthm}
\usepackage{multirow}
\usepackage[titletoc,title]{appendix}
\usepackage{authblk}
\usepackage{setspace}
\usepackage{dsfont}
\usepackage[OT1]{fontenc}
\usepackage{refcount}
\usepackage{booktabs}
\usepackage{arydshln}
\usepackage{bbm}

\usepackage{nameref,zref-xr} 
\zxrsetup{toltxlabel} 

\usepackage{subfig}

\AtEndDocument{\refstepcounter{theorem}\label{finalthm}}
\AtEndDocument{\refstepcounter{equation}\label{finaleq}}

{
  \theoremstyle{definition}
  \newtheorem{assumption}{}
}
{
  \theoremstyle{definition}
  
}
{
  \theoremstyle{definition}
  \newtheorem{example}{}
}
{
\usepackage[utf8]{inputenc}
\usepackage{amssymb,amsmath,amsthm}
\usepackage[margin=1in]{geometry}
\usepackage{parskip}
\usepackage{authblk}
\usepackage{algorithm}
\usepackage{algpseudocode}
\usepackage{arydshln}
\usepackage{array}

\newcolumntype{P}[1]{>{\centering\arraybackslash}p{#1}}

\renewcommand{\P}{\mathsf{P}}
\newcommand{\m}{\mathsf{m}}
\newcommand{\p}{\mathsf{p}}
\newcommand{\q}{\mathsf{q}}

\newcommand{\g}{\mathsf{g}}
\newcommand{\rr}{\mathsf{r}}
\newcommand{\IF}{\mathbb{IF}}

\newcommand{\dd}{\mathsf{d}}

\newcommand{\E}{\mathsf{E}}

\setlength{\parindent}{0pt}

\newcommand{\titlepaper}{Two-step targeted minimum-loss based estimation for non-negative two-part outcomes}

\date{\today}

\author[1]{Nicholas T. Williams}
\author[2]{Richard Liu}
\author[1]{Katherine L. Hoffman}
\author[1]{Sarah Forrest}
\author[1]{Kara E. Rudolph\thanks{KER and ID are both senior authors}}
\author[2]{Iv\'an {D\'iaz}$^*$}

\affil[1]{\small Department of
  Epidemiology, Mailman School of Public Health, Columbia University.}
\affil[2]{\small Division of Biostatistics, Department of Population
  Health, New York University Grossman School of Medicine.}

\begin{document}

\title{\titlepaper}

\maketitle

\begin{abstract}
Non-negative two-part outcomes are defined as outcomes with a density function that have a zero point mass but are otherwise positive. Examples, such as healthcare expenditure and hospital length of stay, are common in healthcare utilization research. Despite the practical relevance of non-negative two-part outcomes, few methods exist to leverage knowledge of their semicontinuity to achieve improved performance in estimating causal effects. In this paper, we develop a nonparametric two-step targeted minimum-loss based estimator (denoted as hTMLE) for non-negative two-part outcomes. We present methods for a general class of interventions, which can accommodate continuous, categorical, and binary exposures. The two-step TMLE uses a targeted estimate of the intensity component of the outcome to produce a targeted estimate of the binary component of the outcome that may improve finite sample efficiency. We demonstrate the efficiency gains achieved by the two-step TMLE with simulated examples and then apply it to a cohort of Medicaid beneficiaries to estimate the effect of chronic pain and physical disability on days' supply of opioids.
\end{abstract}

\section{Introduction}

Non-negative, right-skewed variables with a point mass at zero are commonly encountered in health research (e.g., healthcare expenditures \citep{smith2014marginalized}, hospital length of stay \citep{xie2004method}, daily alcohol consumption \citep{olsen2001two,liu2016analyzing}). These outcomes are often characterized as arising from two distinct processes: the first controlling whether an observation is zero (the \textit{binary} component), and the second controlling the value conditional on the observation not being zero (the \textit{intensity} component). Consider, for example, the days supply of opioids prescribed to a patient. The binary component controls whether or not a patient receives a prescription for opioids, while the intensity component controls the days supply of opioids prescribed conditional on the patient receiving an opioid prescription. 

Approaches to modeling non-negative two-part outcomes include only modeling the binary process or restricting the analysis to observations where the outcome is greater than zero — both of which may lead to erroneous conclusions. Consider a hypothetical scenario where there is no difference in the probability of receiving an opioid prescription between patients with physical disability and those with chronic pain. Nonetheless, among the subset of patients who do receive an opioid prescription, individuals with chronic pain are more likely to be prescribed a greater supply of opioids than those with physical disability. This effect would be overlooked if receiving an opioid prescription or not was the sole outcome considered. Conversely, if we limit the analysis to patients who received an opioid prescription, we are assuming that chronic pain and physical disability have no effect on the likelihood of receiving an opioid prescription; our effect estimate may be biased if this assumption does not hold true. 

Instead of either of the above approaches, the standard approach for modeling non-negative two-part data is with parametric two-part mixture models, often referred to as \textit{hurdle models}, with the binary component modeled using logistic or probit regression and the intensity component modeled using linear regression on the $\log$ scale \citep{duan1983comparison,smith2014marginalized}. This approach results in two sets of covariate coefficients; if the goal is to conduct causal inference on the marginal mean, parametric hurdle models do not yield a simple interpretation of causal effects. In response, \citet{smith2014marginalized} developed a parametric marginalized two-part model that yields more interpretable estimates but still relies on correct parametric specification. More recently, several non-parametric estimators for causal effects with non-negative two-part outcomes have been proposed. \citet{armanBayes} developed a non-parametric Bayesian estimator and demonstrated its improved performance compared to a parametric hurdle model and other non-parametric estimators that do not model the outcome as two parts. \citet{hulingITR} developed a non-parametric estimator for individual treatment rules with semicontinuous outcomes. 

We add to this recent literature by proposing a crossfitted, two-step targeted minimum-loss based estimator (TMLE) for non-negative two-part outcomes. Our approach is non-parametric, doubly-robust, and capable of incorporating data-adaptive estimators of nuisance parameters while achieving $n^{1/2}$ consistency. Improving upon other estimators for causal effects for non-negative two-part outcomes, we define our proposed estimator in terms of general hypothetical interventions on the exposure, allowing effects to be estimated for binary, categorical, and continuous treatments/exposures \citep{haneuse2013estimation,young2014identification,munoz2012population,lmtpJASA}. 

In Section \ref{section2}, we define notation, the target parameter, and its identification result; in Section \ref{section3}, we present the efficiency bound in the non-parametric model; in Section \ref{section4}, we discuss the proposed two-step TMLE, followed by the results of a brief numerical study in Section \ref{section5}. Lastly, as a motivating example, in Section \ref{section6}, we apply our method to estimate the associations between having a physical disability and/or chronic pain condition vs. neither at the time of Medicaid enrollment on the number of days prescribed an opioid during the subsequent 18 months following enrollment among non-elderly, adult Medicaid beneficiaries.

\section{Notation and Definition of the Target Parameter}\label{section2}

Let $O_1, ..., O_n$ denote $n$ i.i.d. observations of observed data $O = (X, T, Y = \Delta S)$, drawn from some distribution $P$, where $X$ denotes a vector of observed covariates that precede a binary, categorical, or continuous exposure $T$, and a non-negative outcome $Y = \Delta S$ such that $\Delta = \mathbbm{1}(Y > 0)$ and $S$ is the value of $Y$ if $Y > 0$. Let $\g(t, X)$ denote $\P(T=t\mid X)$; $\q(t, X)$ denote $\P(\Delta=1|T=t,X)$; and $\m(t, X)$ denote $\E\{Y \mid T=t, X, \Delta = 1\}$. Note that $\E\{Y\mid T, X\} = \E\{Y \mid T, X, \Delta = 1\}\P(\Delta = 1 \mid T, X) + \E\{Y \mid T, X, \Delta = 0\}\P(\Delta = 0 \mid T, X)$ and that $\E\{Y \mid T, X, \Delta = 0\} = 0$. Thus, $\E\{Y\mid T, X\} = \E\{Y \mid T, X, \Delta = 1\}\P(\Delta = 1 \mid T, X) = \m(T, X)\q(T, X)$. We formalize the definition of causal effects using a non-parametric structural equation model (NPSEM): assume the existence of deterministic functions $f_X$, $f_T$, $f_{\Delta}$, and $f_S$ such that $X = f_X(U_X)$; $T = f_T(X, U_T)$; $\Delta = f_{\Delta}(X, T, U_{\Delta})$; $S = f_{S}(X, T, U_{S})$; and $U = (U_X, U_T, U_{\Delta}, U_S)$ is a vector of exogenous variables \citep{pearl2009causality}.

\subsection{General, hypothetical interventions on the exposure}

In what follows, the interventions of interest will be characterized by a function $\dd(t, x, \epsilon)$ that maps $T$, $X$, and potentially a randomizer $\epsilon$ to a new value of $T$. Our target parameter is $\psi = \E\{Y(T^\dd)\}$; the variable $Y(T^{\dd})$ is the counterfactual value of $Y$ that would be observed if the structural equation model for the exposure, $f_T$, was replaced with the output of the user-defined function $\dd$, i.e., we define $T^\dd = \dd(T, X, \epsilon)$ and $Y(T^\dd) = f_{\Delta}(X, T^\dd, U_{\Delta})f_{S}(X, T^\dd, U_{S})$. We focus on defining our target parameter in terms of $\dd(t, x, \epsilon)$ because doing so allows us to estimate a variety of interventions from the causal inference literature for binary, categorical, and continuous valued treatments, including: standard treatment effect measurements, such as the average treatment effect; dynamic treatment regimes; stochastic interventions, such as incremental propensity score interventions; and modified treatment policies \citep{lmtpJASA,hoffman2023introducing}. 

\begin{example}[Static intervention] Let $T$ denote a binary vector, such as receiving a medication, and define $\dd(t, x, \epsilon) = 1$. This intervention characterizes a hypothetical world where all members of the population receive treatment. We refer to this intervention as \textit{static} because the function $\dd$ always returns the same value regardless of the input.
\end{example}

\begin{example}[Dynamic treatment regimes] Let $T$ denote a binary vector, such as receiving a medication, and $X$ a numeric vector, such as a measure of discomfort. For a given value of $\delta$, define
$$\dd(t, x, \epsilon) = \begin{cases}
1 &\text{ if } x > \delta \\
0 &\text{ otherwise.}
\end{cases}$$
Interventions, such as above, where the output of the function $\dd$ depends only on the covariates $X$ are referred to as \textit{dynamic} \citep{hernanWhatIfDynamic,williams2024learning}. 
\end{example}

\begin{example}[Additive shift MTP] Let $T$ denote a numeric vector, such as drug dose. Assume that $T$ has support in the data such that $\P(T \leq u(x) \mid X = x) = 1$. For a value of $\delta$, define
$$\dd(t, x, \epsilon) = \begin{cases}
t + \delta &\text{ if } t + \delta \leq u(x) \\
t &\text{ if } t + \delta > u(x).
\end{cases}$$
This intervention characterizes a hypothetical world where the \textit{natural} value of treatment is increased by a value of $\delta$ wherever an increase is feasible for observations with $u(x)$. This intervention is referred to as a \textit{modified treatment policy} \citep{munoz2012population,haneuse2013estimation,hejazi2023nonparametric,lmtpJASA}.
\end{example}

\begin{example}[Incremental propensity score intervention based on the risk ratio] Let $T$ denote a binary vector, $\epsilon \sim U(0, 1)$, and $\delta$ be an analyst-defined risk ratio between $0$ and $1$. If we were interested in an intervention that decreased the likelihood of receiving treatment, define  
$$\dd(t, x, \epsilon) = \begin{cases}
t &\text{ if } \epsilon < \delta \\
0 &\text{ otherwise.}
\end{cases}$$
In this case, we have $\g^\dd(t, x) = t \delta \g(1, x) + (1 - t) (1 - \delta \g(1, x))$, which leads to a risk ratio of $\g^\dd(1, x)/\g(1, x) =  \delta $ for comparing the propensity score post- vs pre-intervention. 
Conversely, if we were interested in an intervention that increased the likelihood of receiving treatment, define  
$$\dd(t, x, \epsilon) = \begin{cases}
t &\text{ if } \epsilon < \delta \\
1 &\text{ otherwise.}
\end{cases}$$
In this case, we get $\g^\dd(t, x) = t (1 - \delta \g(0, x)) + (1 - t) \delta \g(0, x)$, which implies a risk ratio $\g^\dd(0, x)/\g(0, x) =  \delta $. Because of this interpretation in terms of shifting the propensity score, this type of intervention is referred to as an \textit{incremental propensity score intervention} \citep{kennedy2019nonparametric,wen2023intervention}. 
\end{example}

\subsection{Identification of the causal parameter}

We make the following set of standard assumptions for estimating the causal parameter $\E\{Y(T^\dd)\}$ from observational data:

\begin{assumption}[Positivity]
If $(t, x) \in \text{supp}(T, X)$ then $(\dd(t, x, \epsilon), x)) \in \text{supp}(T, X)$.
\end{assumption}

\begin{assumption}[No Unmeasured Confounding]
$T \perp Y(t)\mid X$ $\forall$ $t$.
\end{assumption}

Assumption 1 states that if there is a unit with observed treatment value $t$ and covariates $x$, there must also be a unit with treatment value $\dd(t, x, \epsilon)$ and covariates $x$. In words, assumption 2 states $X$ contains all common causes of treatment and outcome. Under assumptions 1 and 2, $\E\{Y(T^\dd)\}$ is identified from observed data and equal to

\begin{equation} \label{eq:theta}
\psi = \int \m^\dd(t, x)\q^\dd(t, x)\p(t,x)d\nu(x),
\end{equation}

where $\m^\dd(T, X) = \m(T^\dd, X)$ and $\q^\dd(T, X)=\q(T^\dd, X$).

\section{Efficient Influence Function}\label{section3}

Parametric model-based estimators rely on the assumption that data were generated from a specified parametric distribution. In contrast, semiparametric estimators make few, or possibly, no assumptions about the underlying distribution. Key to constructing semiparametric estimators is the efficient influence function (EIF).

Define $\g(T, X)$ as the density function of $T$ conditional on $X$. We can then specify the density ratio $\rr(t, X) = \frac{\g^\dd(T, X)}{\g(T, X)}$, where $\g^\dd(T, X)$ is the post-intervention density function of $T$. Assume, either, (i) the treatment $T$ is discrete or (ii) the function $\dd$ is piecewise smooth invertible \citep{haneuse2013estimation,lmtpJASA}, and (iii) the function $\dd$ does not depend on the observed distribution $P$. From standard results \citep{munoz2012population,lmtpJASA}, the EIF for $\psi$ is equal to

\begin{equation}\label{eq:eif1}
    D_{\eta,\psi}(O) = \rr(T, X)\{Y - \q(T, X)\m(T,X)\} + \q^\dd(T, X)\m^\dd(T,X) - \psi,
\end{equation}

where we have denoted $\eta=(\m,\q,\rr)$ to represent the set of nuisance parameters. The EIF can alternatively be represented as

\begin{equation}\label{eq:eif2}
    D_{\eta,\psi}(O) = \rr(T, X)\m(T, X)(\Delta - \q(T, X)) + \Delta\rr(T, X) (S - \m(T, X)) + \q^\dd(T, X)\m^\dd(T, X) - \psi.
\end{equation}

The proof is provided in the supplementary materials.

\section{Targeted minimum-loss based estimation}\label{section4}

We propose a two-step targeted minimum-loss based estimator (TMLE), $\hat\psi_{\text{hTMLE}}$, for $\psi$. The targeted minimum-loss based estimator is a substitution estimator that uses estimates of $\hat\eta$ that are constructed to solve the efficient influence function estimating equation \[n^{-1}\sum_{i=1}^nD_{\hat\eta, h\hat\psi_{\text{hTMLE}}}(O_i) = 0.\]We propose using a two-step TMLE since it may improve finite sample efficiency compared to standard TMLE. We use cross-fitting to avoid imposing entropy conditions on the estimates $\hat\eta$. \citep{klaassen1987consistent, Zheng2011, chernozhukov2018double}. Let ${\cal V}_1, \ldots, {\cal V}_J$ denote a random partition of data with indices $i \in \{1, \ldots, n\}$ into $J$ prediction sets of approximately the same size such that 
$\bigcup_{j=1}^J {\cal V}_j = \{1, \ldots, n\}$. For each $j$, the training sample is given by ${\cal T}_j = \{1, \ldots, n\} \setminus {\cal V}_j$. $\hat\eta_{j}$ denotes the estimator of $\eta$, obtained by training the corresponding prediction algorithm using only data in the sample ${\cal T}_j$, and $j(i)$ denotes the index of the validation set which contains observation $i$. The algorithm is as follows:

\begin{enumerate}
    \item Initialize $\hat\eta$. 
    
    The EIF $D_{\eta\psi}(O)$ is a function of the nuisance parameters $\eta = \{\rr, \q, \m\}$, all of which may be estimated by any appropriate regression method in the statistics and machine learning literature. For example, $\m^\dd$ may be estimated by regressing $Y_i$ on $T_i$ and $X_i$ on the subset of observations where $Y_i > 0$ and generating predictions for all observations with $T_i$ set to the output of $\dd(T_i, X_i, \epsilon)$. We use density ratio estimation by classification to estimate $\hat\rr$ \citep{Qin98,cheng2004semiparametric,lmtpJASA}.
    
    \item Fit the tilting model:
    
    \begin{equation}
    \operatorname{logit}\hat\m^{\hat\epsilon_{\m}}(T_i,X_i) = \hat\epsilon_\m + \operatorname{logit}\hat\m(T_i, X_i).
    \end{equation}

    Fit an intercept-only logistic regression of  $Y_i$ with offset equal to $\text{logit } \hat\m(T_i,X_i)$ and weights $\hat\rr_i$ among the subset of observations where $Y_i > 0$. Note that $Y_i$ must first be scaled to the interval $(0, 1)$ before fitting this model.

    \item Update $\hat\m(T_i, X_i) = \m^{\hat\epsilon_{\m}}(T_i, X_i)$.
    \item Fit the tilting model: 

    \begin{equation}
    \operatorname{logit}\hat\q^{\hat\epsilon_{\q}}(T_i,X_i) = \hat\epsilon_\q + \operatorname{logit}\hat\q(T_i, X_i).
    \end{equation}

    Using logistic regression, regress $\mathbbm{1}(Y_i > 0)$ on $\operatorname{logit} \hat\q(T_i,X_i)$ and weights $\hat\rr_i(T_i, X_i) \times \hat\m^{\hat\epsilon_{\m}}(T_i, X_i)$.

    \item Update $\hat\q(T_i, X_i) = \hat\q^{\hat\epsilon_{\q}}(T_i, X_i)$. 
    \item The estimate of $\hat\psi_{\text{hTMLE}}$ is equal to $\frac{1}{n} \sum_{i=1}^n \hat\q^{\dd, \hat\epsilon_{\q}}(T_i, X_i)\hat\m^{\dd,\hat\epsilon_{\m}}(T_i,X_i)$.

    \end{enumerate}

The variance of $\hat\psi_{\text{hTMLE}}$ can be calculated as the empirical variance of Equation \ref{eq:eif1}. However, in finite samples, using the empirical variance of the estimated influence function as a variance estimator for substitution-based estimators, such as $\psi_{\text{hTMLE}}$, often results in anti-conservative variance estimates. We instead estimate the variance of $\psi_{\text{hTMLE}}$ using a non-parametric bootstrap of steps 2-6, such that the variance of $\hat\psi_{\text{hTMLE}}$ is equal to the variance of $b = 1, 2, ..., B$ bootstrapped estimates of $\hat\psi_{\text{hTMLE}}$ \citep{tran2023robust}.

The two-step TMLE is doubly-robust to nuisance parameter model misspecification; it is possible to construct a consistent estimator of $\psi_{\text{hTMLE}}$ if $\hat\rr$, or $(\hat\q, \hat\m)$ is consistently estimated. A software implementation of the algorithm is available for download from \href{https://github.com/mtpverse/hmtp.git}{GitHub} as an \texttt{R} package.

\section{Monte Carlo Unit Testing}\label{section5}

We performed a brief simulation study to illustrate the performance of the two-step TMLE and to ensure correct implementation. We conducted 1000 simulations for samples size $n \in (500, 1000, 5000)$ and compared the performance of the proposed estimator to two other semiparametric estimators commonly used in the causal inference literature: standard TMLE, and augmented inverse probability weight (AIPW); all estimators were estimated using cross-fitting with 10-folds. Estimator performance was evaluated in terms of absolute bias, Monte Carlo variance, mean squared error (MSE), and nominal 95\% confidence interval coverage. We considered the following data generating mechanism:

\begin{table}[H]
\setlength{\tabcolsep}{1pt}
\centering
\footnotesize
\label{tab:dgm}
\begin{tabular}{rl}
$X = (X_1, X_2, X_3, X_4)^\top$ & $\sim N(0, I_4)$ \\ 
$P(T = 1 \mid X)$ & $= \operatorname{logit}^{-1}(\beta_p - X_1 + 0.5 X_2 - 0.25 X_3 - 0.1 X_4)$ \\
$P(\Delta = 1 \mid T,X)$ & $=\operatorname{logit}^{-1}(\alpha_\Delta -0.4 X_1^2 + 0.1 X_2 + 0.8 X_3 - 0.3 X_4 + 2T)$ \\
$U$ & $\sim \operatorname{exp}(\lambda = 1)$\\
$S \mid T, X$ & $\sim e^{(0.1 + 0.2 X_1 + 0.4 X_2 + 0.8 X_3 + 0.3 X_4 + 2T)} + U$ \\
$Y$ & $= \Delta S$ \\
\end{tabular}
\label{tab:sim}
\end{table}

where $\beta_p$ controls positivity and $\alpha_\Delta$ controls the frequency of zeroes. Data were generated under two parameterizations of $\beta_p$: -3, severe positivity violations; and 0, minimal positivity violations; as well as two parameterizations of $\alpha_{\Delta}$: $-2$ and $0$. The target parameter was defined as the mean counterfactual value of $Y$ if $T=1$: $\dd(t, X, \epsilon) = 1$; the true value of $\psi$ under this intervention is approximately $11.99$ when $\alpha_{\Delta} = 0$ and $7.45$ when $\alpha_{\Delta} = -2$. We emphasize that all estimators evaluated the same target parameter. Nuisance parameters were fit using an ensemble of a main-effects GLM and multivariate adaptive regression splines \citep{earth} using the super learner algorithm \citep{vanderLaanPolleyHubbard07}. The super learner optimally combines predictions from a library of candidate algorithms. 

\subsection{Results}

Simulation results are shown in Tables \ref{tab:sim} and \ref{tab:sim2}. In general, estimator performance improved as sample size increased. Regardless of the degree of severity of the practical positivity violations, TMLE and AIPW remained more biased than hTMLE. The discrepancy in bias becomes more pronounced with increasing severity of practical positivity violations. This is expected and it illustrates the importance of using separate models for the point-mass and for the continuous part of the model. 

The two-step TMLE outperformed TMLE and AIPW in terms of MSE and variance across all scenarios. When data were generated with severe positivity violations, all estimators exhibited larger variance, MSE, and suffered from less than nominal confidence interval coverage. However, only the two-step TMLE recovered in terms of nominal confidence interval coverage with increasing sample size; for example, when $n = 5000$ and $\alpha_{\Delta} = 0$, the absolute bias of the two-step TMLE was $0.19$ and confidence interval coverage was $0.95$, compared to the standard TMLE which has an absolute bias of $0.32$ and $0.86$ confidence interval coverage. 

The two-step TMLE also exhibited efficiency gains. When $n = 5000$, $\alpha_{\Delta} = 0$, and $\beta_p = 0$, the variance of the two-step TMLE was 22\% smaller than the the variance of AIPW and TMLE; when $\beta_p = -3$, the variance was of the two-step TMLE was 55\% smaller than the variance of AIPW and 57\% smaller than the variance of TMLE. 

\begin{table}[h]
\caption{Simulation results for the two-step TMLE (hTMLE), TMLE, and augmented inverse probability weighting (AIPW) estimators when $\alpha_{\Delta} = 0$. The target parameter is the expected value under a hypothetical intervention where all observations receive treatment, $\dd(t, x, \epsilon) = 1$; $\psi \approx 11.99$.}
\centering
\footnotesize
\begin{tabular}[]{cccccccc}
\toprule
\multirow{2}{*}{$n$} & \multirow{2}{*}{} & \multicolumn{2}{c}{hTMLE} & \multicolumn{2}{c}{TMLE} & \multicolumn{2}{c}{AIPW} \\
\cmidrule(r){3-4} \cmidrule(lr){5-6} \cmidrule(lr){7-8}
& & $\beta_p = 0$ & $\beta_p = -3$ & $\beta_p = 0$ & $\beta_p = -3$ & $\beta_p = 0$ & $\beta_p = -3$ \\
\midrule
\multirow{4}{*}{500} & $|\text{Bias}|$ & 0.05 & 0.01 & 0.20 & 0.41 & 0.09 & 0.12\\
 & Var. & 0.69 & 4.27 & 0.94 & 15.98 & 0.99 & 39.26\\
 & MSE & 0.69 & 4.26 & 0.98 & 16.13 & 1.00 & 39.24\\
 & Coverage & 0.93 & 0.90 & 0.92 & 0.87 & 0.93 & 0.87\\
 \midrule
\multirow{4}{*}{1000} & $|\text{Bias}|$ & 0.07 & 0.00 & 0.18 & 0.39 & 0.08 & 0.12\\
 & Var. & 0.30 & 2.22 & 0.39 & 5.78 & 0.43 & 10.01\\
 & MSE & 0.31 & 2.22 & 0.42 & 5.92 & 0.43 & 10.01\\
 & Coverage & 0.95 & 0.92 & 0.94 & 0.87 & 0.95 & 0.87\\
 \midrule
\multirow{4}{*}{5000} & $|\text{Bias}|$ & 0.04 & 0.19 & 0.10 & 0.32 & 0.06 & 0.34\\
 & Var. & 0.07 & 0.46 & 0.09 & 1.02 & 0.09 & 1.07\\
 & MSE & 0.07 & 0.49 & 0.10 & 1.12 & 0.10 & 1.18\\
 & Coverage & 0.94 & 0.95 & 0.93 & 0.86 & 0.94 & 0.86\\
\bottomrule
\end{tabular}
\label{tab:sim}
\end{table}

\begin{table}[h]
\caption{Simulation results for the two-step TMLE (hTMLE), TMLE, and augmented inverse probability weighting (AIPW) estimators when $\alpha_{\Delta} = -2$. The target parameter is the expected value under a hypothetical intervention where all observations receive treatment, $\dd(t, x, \epsilon) = 1$; $\psi \approx 7.45$.}
\centering
\footnotesize
\begin{tabular}[]{cccccccc}
\toprule
\multirow{2}{*}{$n$} & \multirow{2}{*}{} & \multicolumn{2}{c}{hTMLE} & \multicolumn{2}{c}{TMLE} & \multicolumn{2}{c}{AIPW} \\
\cmidrule(r){3-4} \cmidrule(lr){5-6} \cmidrule(lr){7-8}
& & $\beta_p = 0$ & $\beta_p = -3$ & $\beta_p = 0$ & $\beta_p = -3$ & $\beta_p = 0$ & $\beta_p = -3$ \\
\midrule
\multirow{4}{*}{500} & $|\text{Bias}|$ & 0.16 & 0.64 & 0.03 & 0.00 & 0.06 & 0.32\\
 & Var. & 0.81 & 3.85 & 1.18 & 10.07 & 25.38 & 74.19\\
 & MSE & 0.84 & 4.25 & 1.18 & 10.06 & 25.36 & 74.21\\
 & Coverage & 0.93 & 0.89 & 0.92 & 0.85 & 0.92 & 0.82\\
 \midrule
\multirow{4}{*}{1000} & $|\text{Bias}|$ & 0.07 & 0.52 & 0.08 & 0.05 & 0.21 & 0.12\\
 & Var.& 0.41 & 2.04 & 0.77 & 4.50 & 8.16 & 11.13\\
 & MSE & 0.41 & 2.30 & 0.77 & 4.50 & 8.20 & 11.13\\
 & Coverage & 0.93 & 0.88 & 0.93 & 0.86 & 0.93 & 0.83\\
 \midrule
\multirow{4}{*}{5000} & $|\text{Bias}|$ & 0.02 & 0.08 & 0.08 & 0.30 & 0.08 & 0.24\\
 & Var. & 0.07 & 0.39 & 0.09 & 0.89 & 0.10 & 1.13\\
 & MSE & 0.07 & 0.39 & 0.10 & 0.98 & 0.10 & 1.19\\
 & Coverage & 0.94 & 0.90 & 0.94 & 0.84 & 0.93 & 0.83\\
\bottomrule
\end{tabular}
\label{tab:sim2}
\end{table}

\section{Illustrative application}\label{section6}

We applied the proposed estimator to estimate the association between physical health status at the time of Medicaid enrollment and the number of days of prescribed opioids among a cohort of 2,440,932 beneficiaries. Physical health status at the time of Medicaid enrollment was operationalized as four mutually exclusive categories defined during a 6-month look-back period: (1) physical disability and co-occurring chronic pain (n = 6,717), (2) physical disability without chronic pain (n = 50,999), (3) chronic pain without physical disability (n = 77,696), and (4) neither physical disability nor chronic pain (n = 2,305,520). Physical disability was identified based on meeting criteria for Medicaid enrollment related to disability status and excluding alternative qualifying disabilities \citep{hoffman2023independent}. Chronic pain status was identified by ICD-10 codes for non-cancer diagnoses typically associated with chronic pain. A beneficiary met the criteria if they had two codes for the same condition occurring during the 6-month look-back period with a minimum separation of 90 days. We estimated effects for the following four interventions: 
\begin{align*}
   \dd_1(t, x, \epsilon) &= \text{physical disability and co-occuring chronic pain}\\
   \dd_2(t, x, \epsilon) &= \text{chronic pain without physical disability}\\
   \dd_3(t, x, \epsilon) &= \text{physical disability without chronic pain}\\
   \dd_4(t, x, \epsilon) &= \text{neither physical disability nor chronic pain}
\end{align*}

A rolling window of 6-months days supply of opioids was calculated by summing the days supply that providers marked for the opioid prescription from the Medicaid pharmacy files. The average of all 6-month rolling windows throughout the study duration was taken for the outcome of average 6-month days supply of opioids prescription. Days of opioids prescribed can be decomposed into two distinct components: whether or not an observation was prescribed any opioids and the number of days opioids were prescribed among those with a prescription.

We considered the following variables as potential confounders: age in years at the time of Medicaid enrollment, sex, race/ethnicity, English as the primary language, marriage/partnership status, household size, veteran status, income likely $>$133\% of the Federal Poverty Level, any inpatient or outpatient diagnosis of bipolar disorder, any anxiety disorder, attention deficit hyperactivity disorder (ADHD), any depressive disorder, or other mental disorders (e.g., anorexia, personality disorders).

All Nuisance parameters were crossfit with 2-folds using the H2O AutoML algorithm with a max runtime of 5 minutes per fold per nuisance parameter \citep{H2OAutoML20, h2oR}. The best performing candidate algorithm was chosen using cross-validated logloss or cross-validated root mean squared error as appropriate for the nuisance parameter. Candidate algorithms included random forests, gradient boosting, generalized linear models with regularization, neural networks, and stacked ensembles of these learners.

Results are presented in Table \ref{tab:results}. After adjusting for baseline confounders and informative right censoring, the estimated average number of days of opioids prescribed was 3.72 (95\% CI: 3.69 to 3.75) among individuals without a physical disability and co-occurring chronic pain, 9.78 (95\% CI: 9.42 to 10.14) among individuals with physical disability and without co-occurring chronic pain, 23.97 (95\% CI: 23.53 to 24.41) among individuals with chronic pain and without co-occurring physical disability, and 44.59 (95\% CI: 41.59 to 47.54) among individuals with physical disability and co-occurring chronic pain. Co-occurring chronic pain and physical disability resulted in 40.84 (95\% CI: 37.87 to 43.82) more days of prescribed opioids compared to having neither chronic pain nor disability. Chronic pain alone resulted in a 20.26 (95\% CI: 19.82 to 20.7) increase in the number of days of prescribed opioids compared to neither, while physical disability alone resulted in a 6.06 (95\% CI: 5.7 to 6.42) increase in the number of days of prescribed opioids compared to neither.

Estimates were similar between hTMLE and standard TMLE, except for the estimated average number of days of opioids prescribed among individuals with physical disability and co-occurring chronic pain. It's important to note that the data do exhibit practical positivity violations. For example, conditional on baseline covariates, the minimum estimated probability of having co-occurring physical disability and chronic pain was 0.0000174. This may partially explain the discrepancy in results between the hTMLE algorithm and standard TMLE.

\begin{table}[h]
\footnotesize
\centering
\caption{Estimates of the expected number of days prescribed opioids for each physical health status category, and estimates of the average causal effects comparing physical disability and co-occurring chronic pain, only chronic pain, and only physical disability to having neither chronic pain or physical disability on average number of days prescribed opioids.}
\begin{tabular}{lcP{8mm}cP{8mm}c}
\toprule
\multicolumn{2}{c}{} & \multicolumn{2}{c}{hTMLE} & \multicolumn{2}{c}{TMLE} \\
\cmidrule(lr){3-4} \cmidrule(lr){5-6}
 & Unadjusted & $\hat{\psi}$ & 95\% CI & $\hat{\psi}$ & 95\% CI \\ 
\midrule
\multicolumn{6}{l}{Population estimate} \\
\midrule
$\dd_1(t, x, \epsilon)$ & $67.43$ & $44.56$ & $41.59$ - $47.54$ & $57.37$ & $56.55$ - $58.19$ \\ 
$\dd_2(t, x, \epsilon)$   & $28.21$ & $23.97$ & $23.53$ - $24.41$ & $24.40$ & $23.94$ - $24.86$ \\ 
$\dd_3(t, x, \epsilon)$   & $14.39$ & $9.78$  & $9.42$  - $10.14$  & $10.30$ & $ 9.94$ - $10.67$ \\ 
$\dd_4(t, x, \epsilon)$   & $3.61$  & $3.72$  & $3.69$  - $3.75$   & $ 3.71$ & $ 3.69$ - $ 3.74$ \\
\midrule
\multicolumn{6}{l}{Average Causal Effect} \\
\midrule
$\dd_1(t, x, \epsilon) - \dd_4(t, x, \epsilon)$ & $63.82$ & $40.84$ & $37.87$ - $43.82$ & $53.65$ & $52.83$ - $54.47$ \\
$\dd_2(t, x, \epsilon) - \dd_4(t, x, \epsilon)$ & $24.60$   & $20.26$ & $19.82$ - $20.70$ & $20.68$ & $20.22$ - $21.15$ \\
$\dd_3(t, x, \epsilon) - \dd_4(t, x, \epsilon)$ & $10.78$   & $6.06$  & $5.70$  - $6.42$   & $6.59$  & $6.22$ - $6.95$   \\
\bottomrule
\end{tabular}
\label{tab:results}
\end{table}

\section{Conclusion}

We proposed a crossfitted two-step TMLE for non-negative two-part outcomes. Our method is non-parametric and capable of incorporating highly flexible data-adaptive algorithms while remaining $n^{1/2}$ consistent. The target parameter was defined in terms of modified treatment policies, allowing effects to be estimated for binary, categorical, and continuous exposures. We demonstrated that the two-step TMLE can improve finite sample performance compared to other commonly used nonparametric estimators when the goal is to estimate the causal effect of an exposure on a non-negative two-part outcome, and that this increase in performance may be more pronounced when there are violations of the positivity assumption.

We applied the hTMLE algorithm to a cohort of Medicaid beneficiaries to estimate the association between physical health status at the time of Medicaid enrollment and the number of days of prescribed opioids. We found that co-occurring physical disability and chronic pain, physical disability alone, and chronic pain alone were all associated with an increase in the number of days of prescribed opioids. Future improvements could include extending the hTMLE algorithm to the longitudinal setting.

\newpage

\bibliographystyle{unsrtnat}
\bibliography{ref}

\appendix
\section{Alternate representation of the EIF}

$\IF(\psi_1 \psi_2) = \IF(\psi1) \psi_2 + \psi_1 \IF(\psi_2)$

Define $P(\Delta = 1 \mid a, w) = \q(a, w)$ and $\E(Y \mid \Delta = 1, a, w) = \m(a, w)$

\begin{flalign*}
\IF &= \IF \bigg[\sum_w \q(a, w) \m(a, w) \p(w) \bigg] &&\\
&= \sum_w \bigg \{ \IF \big[ \q(a, w) \big] \m(a, w) \p(w) + \q(a, w) \IF \big[ \m(a, w) \big] \p(w) + \q(a, w) \m(a, w) \IF \big[ \p(w) \big] \bigg \} &&\\
&= \frac{\mathbbm{1}(A = a)}{\g(a \mid w)} (\Delta - \q(a, w)) \m(a, w) + \q(a, w) \frac{\Delta \mathbbm{1}(A = a)}{P(\Delta = 1, A = a \mid w)} (Y - \m(a, w)) + \q(a, w) \m(a, w) &&\\
&= \frac{\mathbbm{1}(A = a) \m(a, w)}{\g(a \mid w)}(\Delta - \q(a, w)) + \frac{\Delta \mathbbm{1}(A = a)}{\g(a \mid w)}(Y - \m(a, w)) + \q(a, w)\m(a, w)
\end{flalign*}

\begin{align*}
    \q(a, w) \frac{\mathbbm{1}(\Delta = 1, A = a)}{P(\Delta = 1, A = a \mid w)} &= \mathbbm{1}(\Delta = 1, A = a) \frac{\frac{P(\Delta = 1, A = a, w)}{P(A = a, w)}}{\frac{P(\Delta = 1, A = a, w)}{P(w)}} \\
    &= \mathbbm{1}(\Delta = 1, A = a) \frac{P(\Delta = 1, A = a, w) P(w)}{P(\Delta = 1, A = a, w)P(A = a, w)} \\
    &= \mathbbm{1}(\Delta = 1, A = a) \frac{P(w)}{P(A = a, w)} \\
    &= \mathbbm{1}(\Delta = 1, A = a) \frac{P(w)}{P(A = a \mid w) P(w)} \\
    &= \frac{\mathbbm{1}(\Delta = 1, A = a)}{P(A = a | w)}
\end{align*}

\end{document}